 \documentstyle[preprint,aps]{revtex}

\begin{document}
\draft

\title{ Rotons and Quantum Evaporation from Superfluid $^4$He }

\author{    F. Dalfovo$^1$, A. Fracchetti$^1$, A. Lastri$^1$,
            L. Pitaevskii$^{1,2,3}$,  and S. Stringari$^1$}

\address{ $^1$ Dipartimento di Fisica, Universit\`a di Trento, and INFM,
          38050 Povo, Italy }
\address{ $^2$ Department of Physics, TECHNION, Haifa 32000, Israel }
\address{ $^3$ Kapitza Institute for Physical Problems, 117334 Moscow,
          Russia}
\date{May 24, 1995}
\maketitle

\begin{abstract}
The probability of  evaporation induced by $R^+$ and $R^-$
rotons at the surface of superfluid helium is calculated using time
dependent density functional theory. We consider excitation energies
and incident angles such that  phonons do not take part in the
scattering process. We predict sizable evaporation rates, which
originate entirely from quantum effects. Results for the atomic
reflectivity and for the probability of the roton change-mode
reflection are also presented.
\end{abstract}


\narrowtext

Quantum evaporation occurs in superfluid $^4$He when a high-energy
phonon or roton propagates to the surface where it annihilates and an atom
is ejected  in the free space (see, for example, Ref.~\cite{Wya91}). This
phenomenon is  especially interesting  because of the peculiar dispersion
law  exhibited by  rotons.

Despite the significant experimental
\cite{Joh66,Bal78,Bai83,Wya84,Wya90,Bad94,Ens94}
and theoretical \cite{Wid69,And69,Col72,Car76,Mar92,Mul92}
efforts made  in the last years, the fundamental mechanisms
underlaying the phenomenon  of  quantum evaporation are not yet understood.
The experiments by Wyatt and co-workers
\cite{Wya91,Wya84,Wya90}  have  revealed  that the main process
is a one to one process (one excitation to one atom). This behavior
is confirmed by the separate conservation of the energy and of the
momentum parallel to the surface in the evaporation process. In
contrast measurements of atom condensation \cite{Edw82}
have instead pointed out the importance of non linear processes
associated  with the excitation of ripplons.

The theoretical studies have not yet provided a clear and consistent
picture of quantum evaporation. The reason is that it is very difficult to
develop a reliable description of this phenomenon on a microscopic basis.
In fact a good theory should be able to account for several effects:
\begin{itemize}
\item (i)  a correct description of the structure of the free surface, as
well as of the  elementary excitations of the system;
\item (ii) a  quantum description of the scattering processes involving
the elementary excitations at the surface;
\item (iii)  the inclusion of inelastic channels (multi-phonons,
multi-ripplons).
\end{itemize}

A useful discussion concerning the role of quantum effects has been recently
made in Refs.~\cite{Mar92,Mul92} where it has been  pointed out that, due
to the peculiar form of the maxon-roton dispersion exhibited by superfluid
helium, there are severe  constraints on the structure of the classical orbits
associated with  the elementary excitations when they cross the interface.
In particular one  finds that only phonons and rotons above the maxon energy
(about $14$ K) can  give  rise to evaporation.  Vice-versa, the theory
of classical orbits predicts no evaporation from rotons with
energy smaller than the maxon  energy, because of the  occurrence
of a barrier at the interface. The experimental evidence \cite{Wya91,Wya84}
for   quantum evaporation induced by $R^+$ rotons, of all wave vectors,
is  consequently  an important proof of the crucial
role played by quantum effects.  The quantum  states associated with the
above classical orbits (WKB states) have been also used to carry
out a perturbative description of the scattering  process \cite{Mul92}.
However,  the perturbative approach is not easily justifiable in this
context.
The above discussion has also an important implication on the reflection
coefficient. In fact, according to the classical picture of
Ref.~\cite{Mar92},  atoms travelling
at incidence angles sufficiently large, so that they lie in the shaded
part of the  energy diagram  in  Fig.~\ref{fig:shaded}, are  reflected with
unit probability. (Note that in Fig.~\ref{fig:shaded} and in the following,
$q_x$ is the component of wave vector parallel to the surface, while $z$ is
the orthogonal coordinate). This behavior is contradicted by the
experiments of Ref.~\cite{Edw82}, which instead indicate full condensation
also in that region. Actually experimental data show significant reflection
only when  the perpendicular component $q_z$ of the atom wave vector
is close to zero \cite{Clo92}.

The  purpose of this Letter is to provide a first calculation of the
evaporation  rates using a many body  approach accounting for both the
requirements (i) and (ii) discussed above.  We have chosen the values of
$q_x$  and of the energy of  elementary excitations in such a way that
phonons do not take part in the  scattering process; this corresponds to
considering the shaded region in  Fig.~\ref{fig:shaded}.  According
to classical theory this region is characterized  by full  reflection and
no evaporation. It is consequently an ideal region to test the  importance of
quantum effects.

Our approach is based on the study of a slab of liquid $^4$He confined
in a box of size $L_{box}$, as shown in Fig.~\ref{fig:f}(a). The system
is assumed to be translationally invariant in the $x,y$  direction.
The slab is chosen enough thick ($50\sim70$ \AA) to provide a
quantitatively correct  description  of the behavior of the
semi-infinite medium. The box size  ($L_{box} > 100$ \AA) has been
chosen in order to allow a few  oscillations of the  atom wave function
in the free space.

We have calculated the eigenenergies and eigenfunctions of this system in
the framework of density functional theory (see, for instance,
Refs.~\cite{Las95,Dal95} and references therein). When applied to a Bose
system  this theory describes the fluctuations of the density $\rho$ and of
the  velocity potential $\phi$ according to the expansion
\begin{eqnarray}
\psi({\bf r},t)
& \equiv & \sqrt{\rho({\bf r},t)}\  e^{i\phi({\bf r},t)} \nonumber
\\
& = &  \psi_0(z) + f(z) e^{i (q_x x -  \omega t) } +
g(z) e^{i (q_x x + \omega t) }
\label{eq:psi}
\end{eqnarray}
 where $\psi_0(z) =  \sqrt{\rho_0(z)}$
is fixed by the ground state density of the system and
$f(z)$ and $g(z)$ are real wave functions to be determined, together with
the frequency $\omega$, by solving self-consistently the equations
of motion
\begin{equation}
 \delta \int dt  \int d{\bf r} \left( {\cal H} [\psi^*,\psi] -  \psi^*
i \hbar  {\partial  \over \partial t} \psi \right)  = 0
\label{eq:leastaction}
\end{equation}
linearized with respect to $f$ and $g$. The quantity $E= \int \! d{\bf r}\
{\cal H} [\psi^*,\psi] $
 is the energy  functional of the system (depending on $\rho$ and $\phi$)
which is assumed  to be known. The same functional provides, through the
variational  procedure $\delta (E-\mu N)=0$, the ground state profile
$\rho_0(z)$.  In this work we use the density functional
recently proposed in Ref.~\cite{Dal95}. It
provides a  correct description of the equation of state of
superfluid helium, as well  as of the density profile at the surface.
Furthermore it reproduces the  dispersion law of  phonons,
maxons and rotons. The same theory gives a reliable description of
surface modes (ripplons) both at small and high momenta \cite{Las95}.

Equations (\ref{eq:psi},\ref{eq:leastaction}) have the typical form of the
equations of the random phase approximation (RPA).
In particular they account for  both
particle-hole ($f(z)$)  and hole-particle ($g(z)$) transitions which
are coupled by the equations of motion  (\ref{eq:leastaction}). This
coupling is of crucial importance in order  to treat  the
correlation effects associated with the propagation of  elementary
excitations in an interacting system. The equations of motion have also
a structure formally  identical to the one  of the Bogoliubov
equations for the dilute Bose gas and to the one of the Beliaev
equations for Bose superfluids \cite{Mul92,Bel58}. In the vacuum they
coincide with the Schr\"odinger equation for the free atom wave function
$f(z)$, while $g(z)$ vanishes.

One should note that the  equations of time dependent density functional
(TDDF) theory correspond to fully quantum mechanical equations and
consequently
account for  the interference and tunneling phenomena which are expected
to play a crucial role in the evaporation process. Of course, due to
linearization, they do not include inelastic  processes
associated with multi-phonons or  multi-ripplons. These effects lie beyond
the present  formulation of the theory  which is essentially a mean field
theory. For the  same reason in our theory  phonons have an infinite life
time and cannot decay into multi-phonons as is  instead known to occur
at energies below $10$ K. Despite the absence of inelastic processes, we
think that the  solution of the evaporation problem within linearized
TDDF theory is nevertheless highly instructive.

The solution of the equations (\ref{eq:leastaction}) can be determined
with high precision  working in the slab geometry  discussed above.
The solutions are real and either symmetric or antisymmetric with
respect to the  center of the slab. A typical solution is shown in
Fig.~\ref{fig:f}(a), where we plot the function $f(z)$, for an excitation
at $q_x=0.7$~\AA$^{-1}$ and $\hbar \omega = 11$ K.  The figure shows
the existence of atoms travelling outside  the slab and  of elementary
excitations inside the slab.  The corresponding function $g(z)$, not
shown,  has also an
oscillatory behavior inside the slab, while it vanishes outside consistently
with the fact that the hole-particle components of the wave function
(\ref{eq:psi}) (associated with correlation effects) are absent in the
free atom region. The wave length of the atom wave function
is easily calculated starting from the energy conservation law
$ \hbar \omega  = [\hbar^2 q^2 /(2m) - \mu ]  $
where $q^2= q^2_x + q^2_z$, while $\mu =-7.15K$ is the chemical potential
of helium atoms. One finds  $\lambda_z = 2\pi/q_z =
16.4 $~\AA\ in agreement  with the numerical results shown in
Fig.~\ref{fig:f}(a).

Due to the value of $q_x$ and $\omega$ the
solution shown in Fig.~\ref{fig:f}(a)
cannot   contain   phonon components. This is best illustrated in
Fig.~\ref{fig:f}(b) where we  show the  Fourier transform of the
signal inside the slab.  The signal reveals two distinct peaks,  one
corresponding to a $R^-$ roton with $q_z=1.43$~\AA$^{-1}$
($q=1.59$~\AA$^{-1}$), and the other  to a $R^+$ roton  with
$q_z=2.05$~\AA$^{-1}$ ($q=2.17$~\AA$^{-1}$). In this case the $R^-$ and
$R^+$ rotons propagate at angles of $26^\circ$ and $19^\circ$ relative
to the z-axis, respectively, while the atom outside the slab propagates
at an angle  of $61^\circ$.   The additional oscillatory structure
revealed  by Fig.~\ref{fig:f}(b) originates from the finite size of
the box, smaller than $L_{slab}$,  used to calculate the Fourier transform.
In fact one can well fit the calculated signal starting from a function of
the form
\begin{equation}
 f(z) =  f_+ \cos(q^+_zz) + f_- \cos(q^-_zz)
\label{eq:f}
\end{equation}
inside the slab and $f(z)=0$ outside, as shown in Fig.~\ref{fig:f}(b).
Actually this fitting procedure has
been used in order to extract the values of $f_+$ and $f_-$
needed for the analysis of the evaporation rates.
The same procedure has been used to analyze the function
$g(z)$.

The results for $f$ and $g$ can be used to calculate the current giving  the
number of elementary excitations crossing the unit surface per unit time
through the following equation:
\begin{equation}
{\bf j}_i = {\bf v}_i(\mid f_i\mid^2 - \mid
g_i\mid^2)
\label{eq:j}
\end{equation}
where ($|f_i|^2 - |g_i|^2$) is the density of elementary excitations
and ${\bf v}_i$ is the group velocity of the $i$-th excitation  ($i=a,+,-$).
The structure of the current (\ref{eq:j})  emphasizes a remarkable feature
of the RPA (or Bogoliubov) equations. Note that only for a free atom
Eq.~(\ref{eq:j}) takes the familiar form ${\bf j} = |f|^2 \hbar {\bf q}/m $.
In a correlated  system, like superfluid helium, the current behaves
quite differently. For instance,  in the long wave length
phonon regime ($q \to 0$),  the group velocity coincides with
the sound velocity and $g \simeq f$.

The current (\ref{eq:j}) is used to calculate the (real) amplitudes
$A_i=\sqrt{|j_i^z| } \hbox{sgn} (f_i) $ of the signal relative to the various
components (atom, $R^{\pm}$ rotons) in the scattering process taking place
at the surface. The amplitude $A_a$ of the signal relative to the outcoming
atom can be related to the ones of the incoming atom and $R^{\pm}$ rotons
through the relation
\begin{eqnarray}
- A_{a} e^{- i q^a_z L_a} = & & S_{aa} A_{a} e^{i q_z^a L_a} + i S_{-a} A_{-}
e^{-i q^-_z L} \nonumber \\
& + & i S_{+a} A_{+} e^{i q^+_z L}
\label{eq:scattering}
\end{eqnarray}
where $S_{ij}$ is the scattering matrix, $2L=L_{slab}$ is the slab
thickness and $L_a=(L_{box}-L_{slab})/2$.
Relation (\ref{eq:scattering}) holds for symmetric states;  a similar
relation can be written for antysymmetric states. Notice that the phonon
contribution to  the  scattering  process is absent due the choice made
for $q_x$ and for the  energy.

The matrix elements of the scattering matrix entering
Eq.~(\ref{eq:scattering}) satisfy the relation $S_{ij}=S_{ji}$
which follows from unitarity and time reversal symmetry conditions.
In terms of these matrix elements the evaporation
probabilities $P^+$ and $P^-$ (relative to $R^+$ and $R^-$  rotons) and the
reflection  coefficient $R$ take the form
\begin{equation}
 P^+ = \mid S_{+a}\mid^2 \; ; \; P^- = \mid S_{-a}\mid^2  \; ; \; R = \mid
S_{aa}\mid^2 \; .
\label{eq:unitarity}
\end{equation}
 Furthermore, due to unitarity,  one has
$P^+ + P^- + R =1$ (this is true only if one ignores inelastic channels, as
in the present theory).

In order to extract the  physical  coefficients
$P^+$, $P^-$ and $R$ it is necessary to obtain various  solutions at
the same values of $q_x$ and energy, involving different combinations  of the
atom and roton signals. This has been achieved by varying the thickness  of
the slab and the size of the box. Of particular importance, for our analysis,
is the occurrence of the so called "resonance" states which are characterized
by  the absence of the atom signal ($A_a=0$) outside the slab, due to
destructive interference.  These  states are useful
because, as Eqs.~(\ref{eq:scattering}) and (\ref{eq:unitarity}) clearly
show, they allow one to identify the ratio $P^+/P^-$ with $|A_-/A_+|^2$.
Results are shown in Fig.~\ref{fig:results}(a) as a function of the roton
energy for two values of $q_x$ ($0.7$~\AA$^{-1}$ and $0.8$~\AA$^{-1}$).
The error bars are mainly
due to the fact that the atom wave function is not exactly vanishing, even
for the  best resonant states  resulting from the numerical solution of
the equations of motion. This produces a statistical uncertainty in the
values of the branching ratio.

Clearly the determination of the  reflection coefficient, as well as
of the other elements of the scattering matrix, requires the
analysis of non resonant states. A more detailed description of the procedure
and a systematic discussion of the  results will be presented in a longer
paper. In Fig.~\ref{fig:results}(b) we show our results for the  evaporation
probabilities and for the reflection coefficient as
a function of energy. The numerical uncertainty on these values
is expected to be less than 10\%.  The analysis allows us  to
estimate also  the probability $P^{+-}$ for the roton change-mode reflection,
$R^+ \leftrightarrow R^-$. We obtain $P^{+-} \simeq 0.3$ and $0.2$ at
$\hbar \omega = 10.8$ K and $11.3$ K, respectively.

The  main conclusions emerging from our results are:
\begin{itemize}
\item Quantum effects give rise to sizable  evaporation  rates
of  rotons in the region of energy and angles where   evaporation is not
allowed classically.
\item $R^-$ rotons turn out to be less active in the evaporation process
than $R^+$ rotons. Evaporation from $R^+$ rotons becomes dominant
when the energy increases.
\item The probability for the roton change-mode reflection is sizable
in the energy interval considered and decreases with energy.
\item The atom reflection coefficient is smaller than  $10 \%$ for energy
greater than about $11$ K, and decreases for higher energies.
\end{itemize}
The value of the reflection coefficient below $11$ K is still too large
with respect to
the experimental data, but nevertheless its sizable decrease from
the classical value $R=1$ reveals  the very important role played by
quantum effects.  The remaining discrepancy with experiments is  likely
associated with inelastic processes not accounted for in the present
calculation.

The above results  concern the region of large $q_x$ and large
incident angles, where phonons do not take part in the
process. For normal impact and energy smaller than the roton minimum (but
larger than $7.15$ K), one has the opposite regime, where only phonons
take part in the scattering process. The results of our calculations
in this case give very small values for the atom reflection coefficient, in
agreement with experiments.

\bigskip

We are indebted to C. Carraro and A.F.G. Wyatt for many fruitful discussions.
This work was partially supported by the US Department of Energy Office
of Basic Sciences under contract no. W-31-109-ENG-38.

%

\begin{figure}
\caption{Spectrum of elementary excitations in superfluid $^4$He. Solid
line: phonon-maxon-roton dispersion; dashed line: threshold for atom
evaporation; dot-dashed line: dispersion of the free surface mode;
shaded area: region of roton-atom processes.}
\label{fig:shaded}
\end{figure}

\begin{figure}
\caption{An example of solution $f(z)$ for  $q_x=0.7$\AA$^{-1}$,
$\hbar \omega=11$K, $L_{slab}=62$\AA\ and $L_{box}=140$\AA.
The dashed line in the upper part (a) is the
density profile of the slab, in arbitrary units. In the lower part (b)
the Fourier transform of $f(z)$ inside the slab is also shown. The best
fit with formula (\protect \ref{eq:f}) (dashed line) is practically
indistinguishable  from the  numerical solution (solid line). }
\label{fig:f}
\end{figure}

\begin{figure}
\caption{Ratio of the $P^+$ and $P^-$ evaporation probabilities (top) and
absolute values of the evaporation and reflection probabilities (bottom)
as a function of energy. Triangles, circles and squares correspond to
$P^+$, $P^-$, and $R$, respectively. All values below $11.5$K
are calculated at fixed parallel wave vector $q_x=0.7$~\AA$^{-1}$, the
others at $q_x=0.8$~\AA$^{-1}$. }
\label{fig:results}
\end{figure}


\begin{references}

\bibitem{Wya91} M. Brown and A.F.G. Wyatt, J. Phys.: Condens.
Matter {\bf 2}, 5025 (1990)

\bibitem{Joh66} W.D. Johnston and J. G. King, Phys. Rev. Lett. {\bf 16},
1191 (1966)

\bibitem{Bal78} S. Balibar, J. Buechner, B. Castaing, C. Laroche, and A.
Libchaber, Phys. Rev. B {\bf 18}, 3096 (1978)

\bibitem{Bai83} M. J. Baird, F.R. Hope, and A.F.G. Wyatt, Nature
{\bf 304}, 325 (1983); F.R. Hope, M.J. Baird, and A.F.G. Wyatt,
Phys. Rev. Lett. {\bf 52}, 1528 (1984)

\bibitem{Wya84} A.F.G. Wyatt, Physica {\bf 126B}, 392 (1984)

\bibitem{Wya90} G.M. Wyborn and A.F.G. Wyatt, Phys. Rev. Lett. {\bf 65},
345 (1990)

\bibitem{Bad94} H. Baddar, D.O. Edwards, T.M. Levin, and M.S. Pettersen,
Physica B {\bf 194-196}, 513 (1994)

\bibitem{Ens94} C. Enss, S.R. Bandler, R.E. Lanou, H.J. Maris, T. More,
F.S. Porter, and G.M. Seidel, Physica B {\bf 194-196}, 515 (1994)

\bibitem{Wid69} A. Widom, Phys. Lett. {\bf 29A}, 96 (1969); D.S. Hyman,
M.O. Scully, and A. Widom, Phys. Rev. {\bf 186}, 231 (1969)

\bibitem{And69} P.W. Anderson, Phys. Lett. {\bf 29A}, 563 (1969)

\bibitem{Col72} M.W. Cole, Phys. Rev. Lett. {\bf 28}, 1622 (1972)

\bibitem{Car76} C. Caroli, B. Roulet, and D. Saint-James, Phys. Rev. B
{\bf 13}, 3875, 3884 (1976)

\bibitem{Mar92} H. J. Maris, J. Low Temp. Phys. {\bf 87}, 773 (1992)

\bibitem{Mul92} P.A. Mulheran and J.C. Inkson, Phys. Rev. B {\bf 46},
5454 (1992)

\bibitem{Edw82} D.O. Edwards, Physica {\bf 109} \& {\bf 110}B, 1531
(1982), and  references therein; V.V. Nayak, D.O. Edwards,
and N. Masuhara, Phys. Rev. Lett. {\bf 50} 990 (1983); S. Mukherjee,
D. Candela, D.O. Edwards, and S. Kumar, Jap. J. Appl. Phys. {\bf 26-3},
257 (1987); A.F.G. Wyatt et al., unpublished.

\bibitem{Clo92} The threshold behavior of the reflection coefficient
has been also discussed in D.P. Clogherty and W. Kohn, Phys. Rev. B {\bf 46},
4921 (1992), in the general context of the quantum theory of sticking.

\bibitem{Las95} A. Lastri, F. Dalfovo, L. Pitaevski, and S. Stringari,
J. Low Temp. Phys. {\bf 98}, 227 (1995)

\bibitem{Dal95} F. Dalfovo, A. Lastri, L. Pricaupenko, S. Stringari, and
J. Treiner, Phys. Rev. B, in press

\bibitem{Bel58} S. T. Beliaev, Zh. Eksp. Teor. Fiz. {\bf 34}, 417 (1958)
[Sov. Phys. JETP {\bf 7}, 289 (1958)]

\end{references}
\end{document}